\documentclass[]{pasj01}

\Received{}
\Accepted{}
 
\begin{document} 

\title{ 
The soft X-ray background with Suzaku II:  Supervirial temperature bubbles?}

\author{Hayato \textsc{Sugiyama}\altaffilmark{1}, Masaki \textsc{Ueda}\altaffilmark{1}, Kotaro \textsc{Fukushima}\altaffilmark{1},  Shogo B. \textsc{Kobayashi}\altaffilmark{1}, Noriko Y. \textsc{Yamasaki}\altaffilmark{2}, Kosuke \textsc{Sato}\altaffilmark{3} and Kyoko \textsc{Matsushita}\altaffilmark{1}}
\altaffiltext{1}{Department of Physics, Tokyo University of Science, 1-3 Kagurazaka, Shinjuku-ku, Tokyo 162-8601, Japan}
\altaffiltext{2}{ Institute of Space and Astronautical Science, Japan Aerospace Exploration Agency, 3-1-1 Yoshinodai, Chuo-ku, Sagamihara, Kanagawa 252-5210, Japan}
\altaffiltext{3}{Graduate School of Science and Engineering, Saitama University, 255 Shimo-Okubo, Sakura-ku, Saitama, Saitama 338-8570, Japan} 
\email{matusita@rs.tus.ac.jp}

\maketitle

\begin{abstract}

Observations of the hot X-ray emitting interstellar medium in the Milky Way are important for studying the stellar feedback and understanding the formation and evolution of galaxies.
We present measurements of the soft X-ray background emission for 130  Suzaku
observations at $75^\circ<l < 285^\circ$ and $|b|>15^\circ$.
With the standard soft X-ray background model consisting of the local hot bubble and the Milky Way halo, residual structures remain at 0.7--1 keV in the spectra of some regions.
Adding a collisional-ionization-equilibrium component with a temperature of $\sim$0.8 keV, much higher than the virial temperature of the Milky Way,  significantly reduces the derived C-statistic for 56 out of 130 observations.
The emission measure of the 0.8 keV component varies by more than an order of magnitude:
Assuming the solar abundance, the median value is 3$\times 10^{-4}~ \rm{cm^{-6} pc}$ and the 16th-84th percentile range is (1--8)$\times 10^{-4}~ \rm{cm^{-6} pc}$.
Regions toward the Orion-Eridanus Superbubble, a large cavity extending from the Ori OB1 association, have the highest emission measures of the 0.8 keV component. 
While the scatter is large,  the emission measures tend to be higher toward the lower Galactic latitude. 
We discuss possible biases caused by the solar wind charge exchange, stars, and background groups. The 0.8 keV component is probably heated by
supernovae in the Milky Way disk, possibly related to galactic fountains.

\end{abstract}

\section{Introduction}

Understanding star formation and stellar feedback is essential for studing the formation and evolution of galaxies.
Supernovae (SNe) heat the interstellar medium (ISM). In star-forming regions, stellar wind or multiple SNe can create superbubbles filled with hot plasma (e.g. \cite{Hill2012,Keller2014,SBreview}).
SNe may drive hot gas from the disk, create outflows or galactic fountains,  and eject energy and metals into the intergalactic space \citep{Dekel86,CGM}.
In addition, accreting gas from the intergalactic space  
is expected to be heated to the virial temperature 
by shock waves generated by mass accretion (e.g., \cite{Rees1977}).
We can study these activities with X-ray observations of hot ISM in galaxies.

 Suzaku, Japan's X-ray astronomy satellite, is characterized by its low background, allowing us to make a detailed study of the low-surface brightness X-ray emission from hot gas in the Milky Way. 
 Such hot gas contributes to the soft X-ray background detected below $\sim$ 1 keV.
A spectral component modeled with a collisional-ionization-equilibrium (CIE) component at a temperature of $\sim$ 0.2 keV is thought to originate from an extended, diffuse plasma in our Galaxy, called the Milky Way Halo (MWH) (e.g., \cite{Yoshino09}, \cite{Henley13},  \cite{Nakashima18}). A cavity filled with a CIE component at a temperature of 0.1 keV, thought to be created by SNe surrounding the Sun, is called the local hot bubble (LHB).
In addition, 
to explain the spectra of some regions, \citet{Yoshino09} needed a much higher Ne abundance (Ne/O$\sim$ 3 solar) or an additional higher temperature (0.5--0.9 keV) component.

\citet{Sekiya14a}, \citet{Nakashima18}, and \citet{Gupta2021} also reported the presence of the 0.6-1.0 keV component from some regions observed with {Suzaku}.
\citet{Gupta2022} analyzed Suzaku data from 230 fields and found a bright, supervirial temperature component from the eROSITA bubble \citep{erositabubble} that includes the North Polar Spur/Loop I structure. \color{black} Their analysis also revealed that the emission measure of this supervirial component exhibits a dependence on the galactic longitude, with a significant
increase toward the bubble. \color{black}
\citet{Halo22} also detected two hot gas model components, with temperatures of $\sim$0.17 keV and $\sim$0.7 keV, which are widely distributed across the Milky Way with the HaloSat \color{black} at $|b|>30^\circ$ and studied the dependence on the angular distance from the Galactic center \color{black}.
With eROSITA observations covering 142 square degrees ($220^\circ<l<235^\circ, 20^\circ<b<40^\circ$), \citet{Ponti22} also found a similar spectral component with temperature of 0.4--0.7 keV.

\citet{Ueda22}, hereafter Paper-I,  analyzed {Suzaku} data from 130 observations ($75^\circ < l< 285^\circ, |b| > 15^\circ$)
and reported that adding a hot CIE component with a temperature of $\sim$0.8 keV (hereafter the 0.8 keV component) to the standard background model reproduces the spectra of a significant fraction of the observations.
In addition, they suggested that O\,\emissiontype{VII} He$\alpha$ emissions, possibly from heliospheric solar wind charge exchange (SWCX), sometimes lead to underestimation and overestimation of temperatures and emission measures, respectively, of the MWH component.
From the observations before the end of 2009, or excluding the data around the solar maximum,
the temperature ($\sim 0.2$ keV) of the MWH component is quite uniform.
At $|b|<35^\circ$, its emission measure increases toward the lower Galactic latitude, indicating the presence of a disk-like morphology component. In addition, they suggested that plasma with the virial temperature ($\sim 0.2$ keV) of the Milky Way may fill the halo in nearly hydrostatic equilibrium.

This paper presents the results of the 0.8 keV component derived from the same observations as in Paper-I.
The paper is organized as follows.
In Section 2, we describe the observations and data reduction.
In Section 3, we present spectral fitting results,
\color{black} the spatial distribution of emission measure of the 0.8 keV component, especially dependence on the Galactic latitude, and systematic uncertainties caused by the SWCX\color{black}.
In Section 4, we discuss the results.
This paper uses the solar abundance table from \citet{Lodders03}.
Errors are reported at the 68\% confidence level unless otherwise noted.

\section{Observations and data reduction}

We have analyzed the \color{black} X-ray Imaging Spectrometer (XIS) data \color{black}  of the 130 observations at $75^\circ<\ell<285^\circ$ and $|b|>15^\circ$ with {\it Suzaku} presented in Paper-I.
This excludes regions toward the Galactic center, the eROSITA bubble, and the Galactic disk.
We also excluded the regions around very bright point sources and extended objects, such as galaxy clusters and supernova remnants.  
The XIS consists of four CCD sensors. XIS0, 2, and 3 contain front-illuminated (FI) CCDs and
XIS1 has a back-illuminated (BI) CCD \citep{Koyama2007}. We analyzed four CCD data taken before the loss of XIS2 in November 2006. 
After the loss,  we analyzed the data from the three remaining CCDs.
Data obtained with 3$\times$3 and 5$\times$5 editing modes were merged.
We applied the same filtering and point source removal as in Paper-I.
To decrease the contamination caused by the SWCX, we created light curves and excluded the time ranges where the count rate exceeded 3$\sigma$.
The details were described in Paper-I.

\section{Spectral analysis and results}

We extracted a spectrum over the field of view (FOV) of each XIS detector from each observation.
We used the xisrmfgen ftools task to create the redistribution matrix files (RMFs).
To create ancillary response files, we used the xissimarfgen task assuming uniform emission from a circular region with a radius of 20$'$.
The instrumental background, or non-X-ray background (NXB), was estimated from the Night-Earth database using the xisnxbgen task.
We used XSPEC version 12.10.1f \citep{xspec} to model the NXB-subtracted spectra.
We used the energy ranges of  0.4--7.0 keV  for the BI spectra, and 0.5--7.0 keV 
  for the FI spectra.
We rebinned each spectrum to have at least one count per bin and employed the extended C-statistic \citep{Cash79}.
We used APEC (Astrophysical Plasma Emission Code, \cite{Smith01}, \cite{Foster12}) with AtomDB version 3.0.9 to model a CIE plasma.

\subsection{The standard soft X-ray background model}

\begin{figure*}
\centerline{
  \includegraphics[width=8cm]{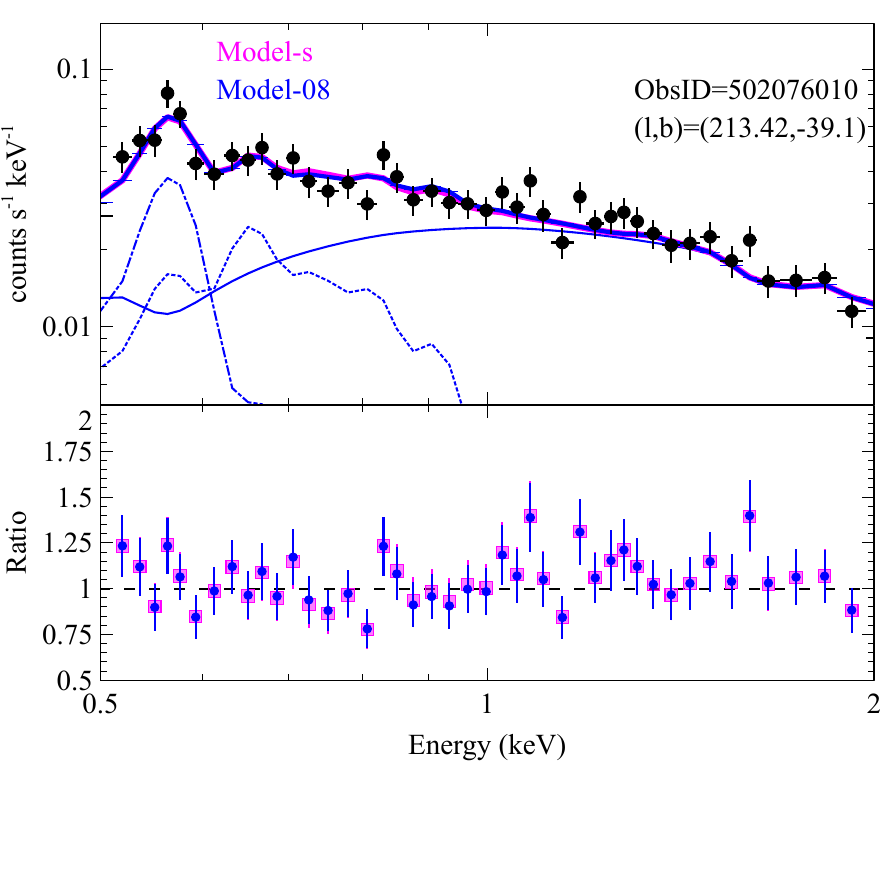} 
 \includegraphics[width=8cm]{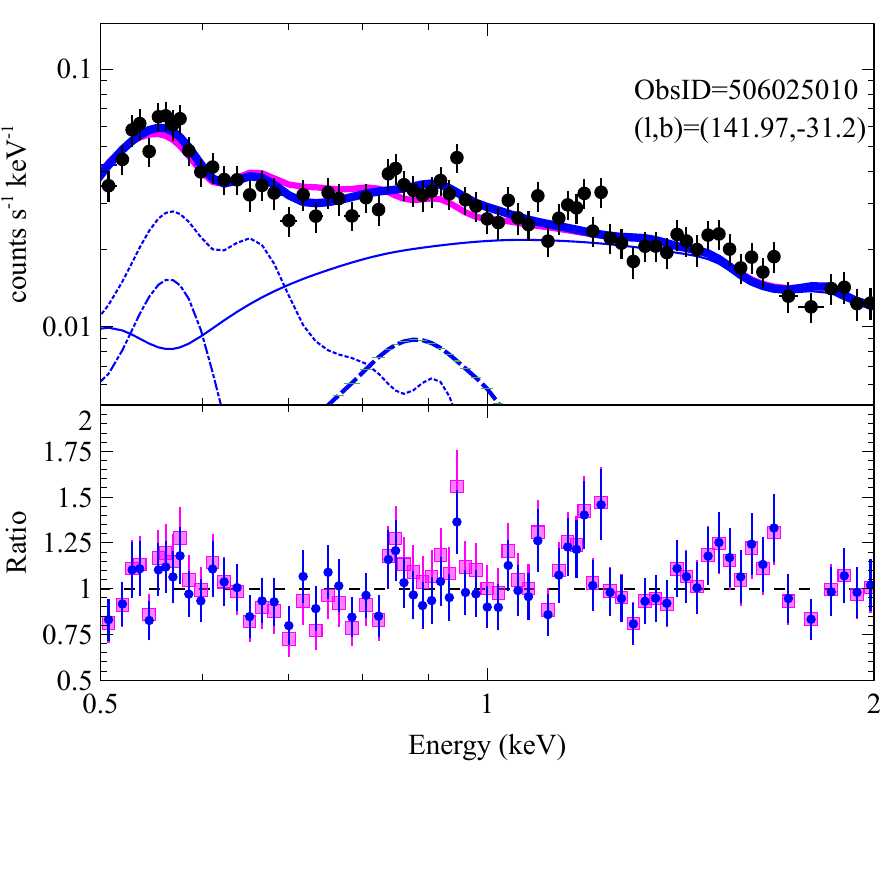} 
}
\centerline{
  \includegraphics[width=8cm]{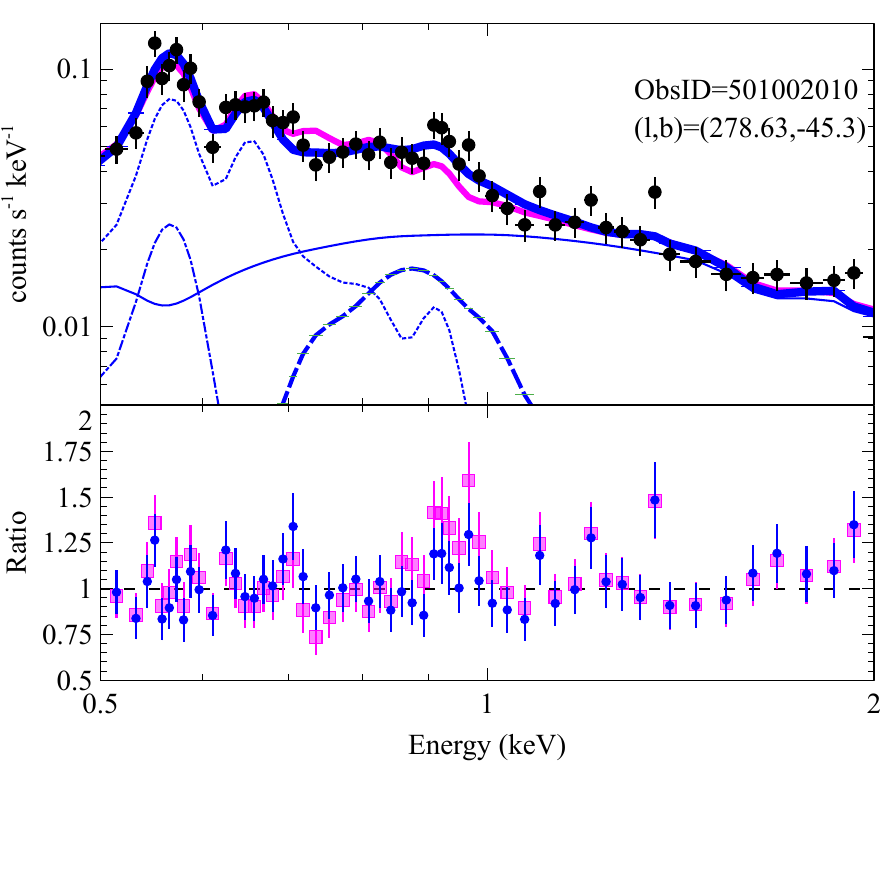} 
 \includegraphics[width=8cm]{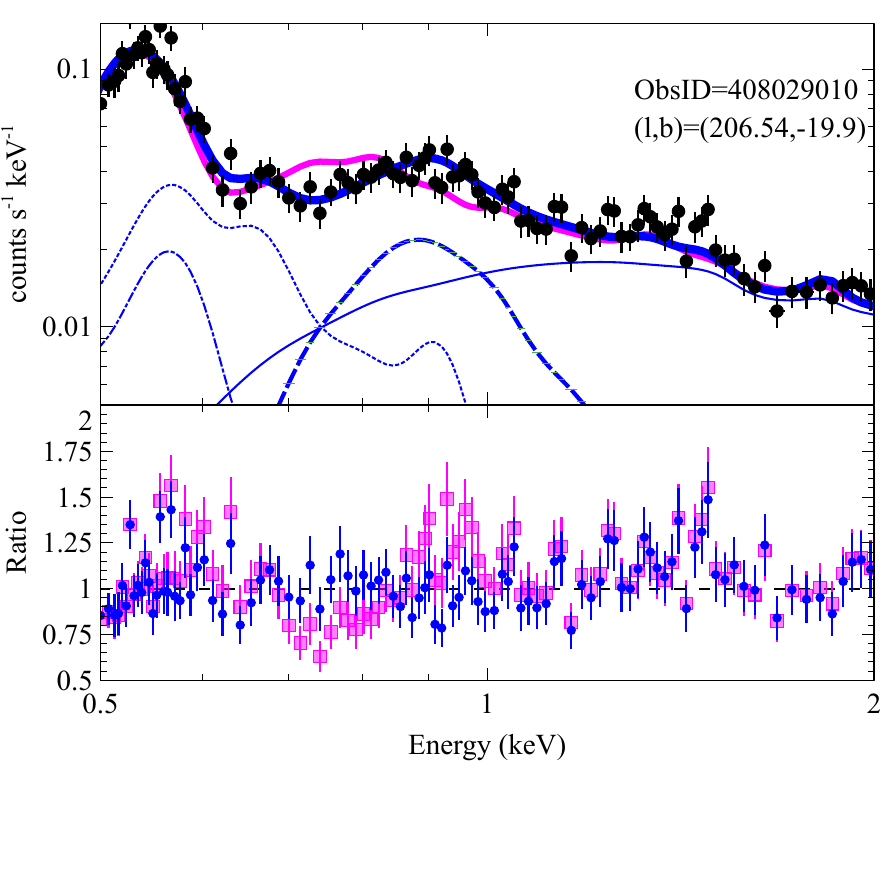} 
}
\caption{(upper panels) Representative XIS1 spectra (ObsID=502076010, 506025010, 501002010, 408029010) fitted with Model-s (thick magenta lines) and  Model-08 (thick blue lines).  
The thick blue dashed lines show the contributions of the 0.8 keV component.
The thin blue solid, dotted, and dot-dashed lines correspond to the contributions of CXB, MWH, and LBH, respectively, to the best-fit Model-08.
(bottom panels) The data-to-model ratio for Model-s (filled magenta squares) and Model-08 (filled blue circles).}	
\label{fig:repspec}
\end{figure*}

We first applied the standard background model, hereafter referred to as Model-s.
This model has four components:
 The cosmic X-ray background (CXB),  two CIE components ("apec" model in XSPEC) to model LHB and MWH, and the O\,\emissiontype {I} K$\alpha$ line for the scattered photons from the sunlit atmosphere of the Earth \citep{Sekiya14OI}.
 We used a power-law model with a photon index of 1.4 to model CXB.
We fixed the temperature of the LHB at 0.1 keV and 
the abundances of LHB and MWH  at 1 solar.
 The CXB and the MWH were subject to photoelectric absorption due to cold gas. We modeled  this absorption using the "phabs" model in the XSPEC spectral fitting tool,  
  with a fixed hydrogen column density at the Galactic value from \citet{Kalberla05}.  
 A Gaussian with fixed central energy of 0.525 keV was used to model the O\,\emissiontype {I} line.
 The normalization of each component was allowed to vary.
 The spectra of the XIS detectors of each observation were fitted simultaneously with Model-s.
 These fits are the same as in Paper-I, where the results for the MWH component were presented.
 In most cases, the temperatures of the MWH component (hereafter $kT_{\rm halo}$) are in the range of 0.15--0.3 keV.
 However, 
as reported by \citet{Yoshino09}, \citet{Sekiya14a}, and \citet{Nakashima18},
some observations show significantly higher MWH temperatures around 0.6--1 keV.
As shown in figure \ref{fig:repspec}, with Model-s,
there are residual structures around 0.7--1 keV in the spectra of some observations.

\subsection{With a supervirial temperature component}\label{sec:model08}

We then added another "apec" component modified by the Galactic absorption to  Model-s and fitted the spectra with this model.
The temperature and abundance of this additional component were fixed at 0.8 keV and 1 solar, respectively.
We will refer to this model as Model-08.
This paper presents the results of the 0.8 keV component with Model-08, while those of the others were presented in Paper-I.
As shown in figure \ref{fig:repspec}, Model-08 can reproduce the spectra well.
Out of 130 observations, 56 of them got $\Delta \rm{C}<-9$. Here, $\Delta$C is the C-statistic for Model-08 subtracted from that for Model-s.
For observations with exposures longer than 60 ks, 35 out of 57 observations have $\Delta\rm{C}<-9$.
Thus, almost half of the observations prefer Model-08 to Model-s,
although some with the longest exposures have $\Delta\rm{C}\sim 0$.
 The emission  measure (=$\int n_{\rm H}n_e ds$, where $s$ is the distance along the line of sight) of the 0.8 keV component (hereafter EM$_{08}$) spans an order of magnitude;  
the median value and the 16th-84th percentile range of the 130 observations (including data with $\Delta \rm{C}\ge-9$) are  $3.8\times 10^{-4}~\rm{cm^{-6}pc}$ and  
  (0.9--7.7)$\times 10^{-4}~\rm{cm^{-6}pc}$, respectively (table \ref{tab:em08median}). 
As reported in Paper-I, the median of $kT_{\rm halo}$ becomes 0.19 keV.

\begin{figure}
\centerline{
   \includegraphics[width=8cm]{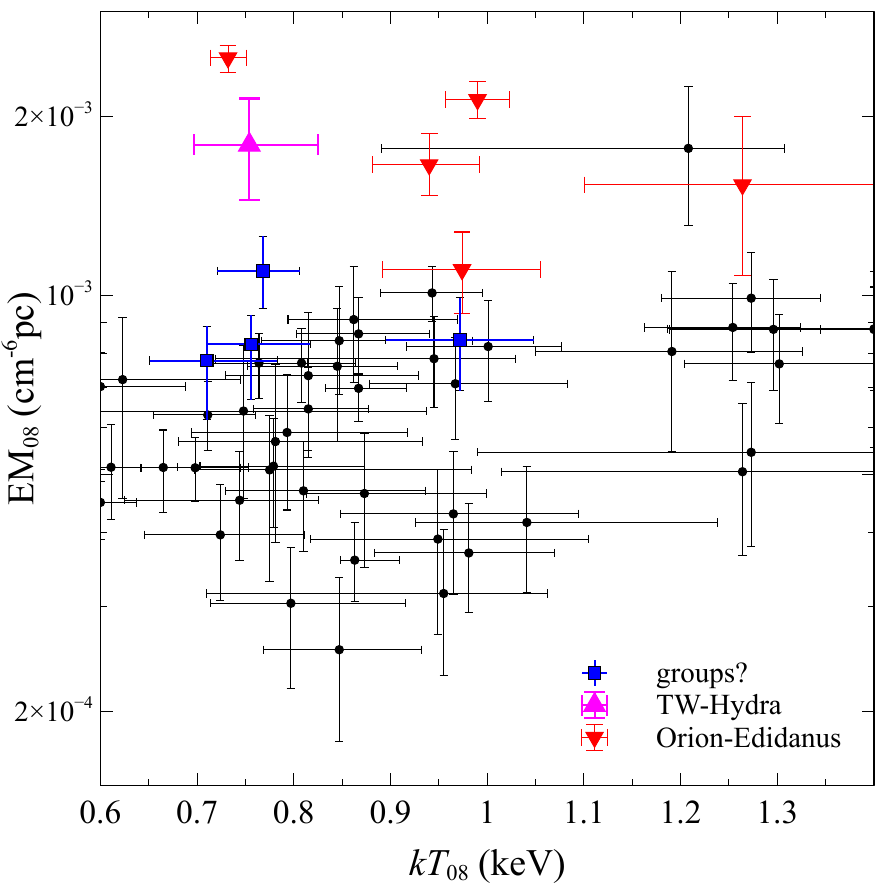}
}
\bigskip
\caption{Temperature of the 0.8 keV component plotted against EM$_{08}$ with Model-08-kTfree for the observation with $\Delta$C$<-9$. Squares indicate possible group candidates (see section \ref{sec:group}), and up and down triangles correspond to TW Hya \color{black}(section \ref{sec:sp})\color{black} and regions toward Orion-Eridunus superbubble \color{black}(sections \ref{sec:sp} and \ref{sec:ori})\color{black}, respectively. 
}
\label{fig:ktem}

\end{figure}

We then allowed the temperature of the 0.8 keV component to vary in the range of 0.6--1.4 keV and re-fitted the spectra of the observations with $\Delta$C$<-9$.
We will refer to this model as Model-08-kTfree.
As shown in figure \ref{fig:ktem}, 
the temperatures of the 0.8 keV component, or $kT_{08}$, are concentrated around 0.8--1 keV.
There are several regions with temperatures around 1.3 keV.
There is no significant correlation between EM$_{08}$ and $kT_{08}$. 
$kT_{08}$ does not depend on the Galactic latitude or longitude.

\begin{table}
            \tbl{Medians and 16th-84th percentile ranges of EM$_{08}$ with Model-08 }{
                      \begin{tabular}{ccccc}\hline 
          selection     & N$^*$ & median & 16th-84th percentile\\
                           &  &  $10^{-4}\rm{cm^{-6}pc}$ & $10^{-4}\rm{cm^{-6}pc}$\\
                 \hline 
                          all & 130 & 3.8 & 0.9--7.7 \\
              $|b|\le 35^\circ$ & 47  & 4.9 & 2.1--9.5 \\
              $35^\circ<|b|\le 50^\circ$ & 42 & 3.8 & 0.8--6.8\\
              $|b|>50^\circ$ & 41 &2.3 & 0.6--4.6\\
	     \hline
            \end{tabular}}\label{tab:em08median}
                \begin{tabnote}
                    \footnotesize
                    \footnotemark[$*$] Number of observations\\           
	      \end{tabnote}
                \end{table}

\color{black}

In our analysis, we adopt the column density of the absorption, $N_{\rm H}$ as the Galactic value of the HI gas.
If the 0.8 keV component is distributed inside the cold gas or coexists with it,  we may overestimate both $N_{\rm H}$ and EM$_{08}$.
Therefore, we refitted the spectra with Model-08, allowing $N_{\rm H}$ of the 0.8 keV component to vary, with an upper limit at the Galactic value. 
For most of the observations, the obtained values of EM$_{08}$ are in agreement with those obtained from Model-08.
In section \ref{sec:spmodel}, we estimate the scale height of the 0.8 keV component, which 
 is likely larger than several tens of pc for the
HI  and H$_2$ disks.  Consequently, the majority of 
 the 0.8 keV component might be located outside the cold gas disks, validating the use of  the "Galactic value" 
 for the spectral modeling.

\subsection{Spatial distribution of the 0.8 keV component}\label{sec:sp}

	\begin{figure*}
				\centerline{\includegraphics[width=15cm]{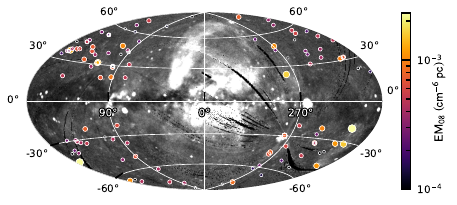}	}
					\caption{The 3/4 keV band image of the ROSAT all-sky survey. The color and size of the marks indicate EM$_{08}$ with Model-08.}
			\label{fig:map08}
\end{figure*}

   Figure \ref{fig:map08} shows EM$_{08}$ on the 3/4 keV band image of the ROSAT all-sky survey.
The brightest EM$_{08}$ regions are concentrated toward the Orion-Eridanus superbubble \citep{Reynolds79, Brown1995} around ($l, b$)$\sim$ (200$^\circ$, -40$^\circ)$. This bubble is a large cavity extending from the Ori OB1 association and spanning 20$^\circ\times 45^\circ$ in the sky.
The observation with ObsID=408029010 ($l$=206.54, $b$=-19.93, bottom-right panel of figure \ref{fig:repspec}), which toward the superbubble,  has the brightest 0.8 keV component.
EM$_{08}$ of other four regions toward the bubble, ObsID=702062010 ($l=211.76, b=-32.06$), ObsID=706013010 ($l=174.83, b=-44.51$), ObsID=407045010 ($l=201.05, b=-31.29$), and ObsID=409029010 ($l=192.85, b=-48.96$) also have  EM$_{08}>10^{-3}~\rm{cm^{-6}pc}$ (figure \ref{fig:ktem}).
EM$_{08}$ around a T Tauri star, TW Hya ($l$=278.66, $b$=22.96, ObsI{\color{black} D}=402089020), is the third brightest among the 130 observations. 
In some cases, nearby regions show significantly different EM$_{08}$.
For example,  the EM$_{08}$ of a region ($l$=213.42, $b$=-39.1, ObsID=502076010) near the Orion-Eridanus superbubble is consistent with zero (the top left panel of figure \ref{fig:repspec}).

\begin{figure*}
    \centerline{
      \includegraphics[width=9cm]{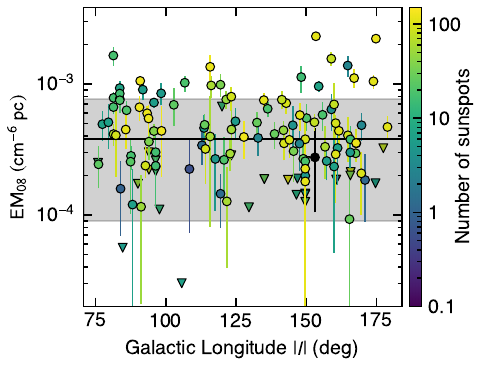} 
   \includegraphics[width=9cm]{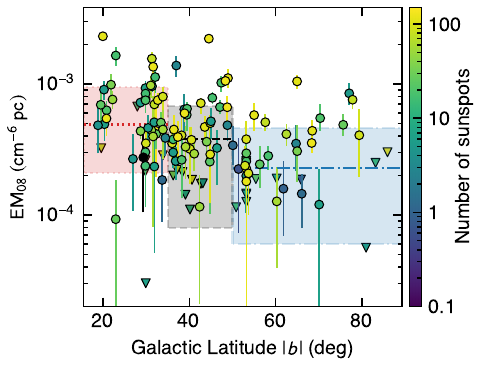} 
    }
\caption{(left) EM$_{08}$ with Model-08 plotted against the  absolute value of the Galactic longitude, $|l|$. The color scale corresponds to the 13-month smoothed sunspot number. The shaded region shows the median and the 16th-84th percentile range for the 130 observations. The triangles show the 1 $\sigma$ upper range. (right)  Same as the left panel but plotted against the absolute value of the Galactic latitude, $|b|$. The shaded regions show the medians and 16th-84th percentile ranges for the observations at $|b|\le 35^\circ$, $35^\circ<|b|\le 50^\circ$, and $|b|>50^\circ$. 
}\label{fig:mwhnormvsbeta}
\end{figure*}

   Figure \ref{fig:mwhnormvsbeta} shows EM$_{08}$ for Model-08 plotted against the absolute value of Galactic longitude, $|l|$, and latitude, $|b|$. 
  Here, $|l|$ is defined as $l$ for $0^\circ\le l <180^\circ$ and $(360^\circ - l)$ for $180^\circ \le l < 360^\circ$. 
  For a given longitude or latitude, there is a significant scatter of an order of magnitude in EM$_{08}$. 
As shown in table \ref{tab:em08median}, 
the median value of the low latitude sample ($|b|<35^\circ$) is a factor of 1.3 and 2 higher than those of the mid-latitude($35^\circ<|b|<50^\circ$) and high-latitude samples ($|b|>50^\circ$), respectively, although the 16th-84th percentile ranges overlap significantly.
There is no significant dependence of EM$_{08}$ on $|l|$.

\subsection{Correlation with the solar activity}
\begin{figure}
 \begin{center}
  \includegraphics[width=8cm]{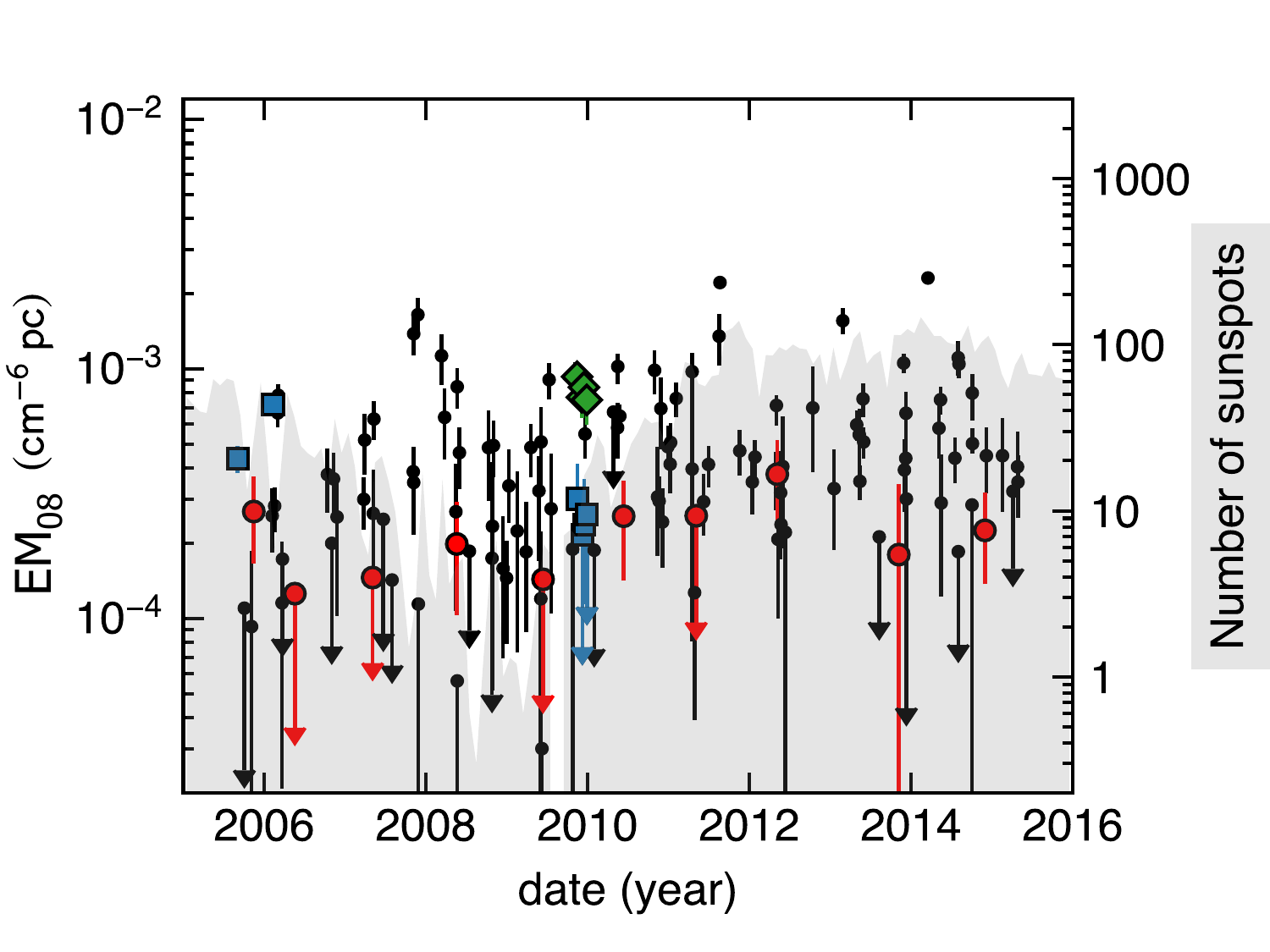}
 \end{center}
 \caption{EM$_{08}$ plotted against the observation date. The 13-month averaged sunspot number,  SSN, is also shown as the gray-shaded area. The filled circles, squares, and diamonds are LH, NEP, and SEP, respectively. The data with the lower triangles show an upper range of 1$\sigma$.}\label{fig:vsdate}
\end{figure}

Our data were obtained from 2005 to 2015, covering almost one solar cycle, including the solar minimum around 2009 and the solar maximum around 2014. 
Therefore, they are suitable for studying the contributions of SWCX.  
\citet{Fujimoto07} found strong O/Ne/Mg line emission from geocoronal SWCX with Suzaku.
In addition to the time-variable O\,\emissiontype{VII}, O\,\emissiontype{VIII}, Ne\,\emissiontype{X}, and Mg\,\emissiontype{XI} emission lines, they also detected features in the 0.75-0.95 keV energy range.
Although we screened the data using lightcurves, 
these SWCX emissions may remain and contribute to the 0.8 keV component.
In addition, it is not easy to filter the heliospheric SWCX emissions, whose time variations are expected to be slow.
In figure \ref{fig:mwhnormvsbeta}, we plot EM$_{08}$ for Model-08 against the Galactic latitude with the 13-month-smoothed sunspot number (hereafter SSN)
from the World Data Center SILSO, Royal Observatory of Belgium,
Brussels (SILSO 2005–2015)
\footnote{SILSO, World Data Center—Sunspot Number and Long-term Solar Observations,
Royal Observatory of Belgium, on-line Sunspot Number catalogue (2005–2015)
$⟨$http://www.sidc.be/SILSO/$⟩$}  in the color scale.
Figure \ref{fig:vsdate} shows EM$_{08}$ plotted against the observation date with SSN.
The scatter in EM$_{08}$ is large, and the correlation with the solar activity is relatively small.
The median value of EM$_{08}$ for the data obtained before the end of 2009 (hereafter 2005-2009 data), $\sim 3 \times 10^{-4}~\rm{cm^{-6} pc}$, is slightly lower than $\sim 4\times 10^{-4}~\rm{cm^{-6} pc}$ for the remaining data (hereafter 2010-2015 data), as shown in table \ref{tab:em08median2}, although the 16th-84th percentile ranges overlap significantly.

\begin{figure}
 \begin{center}
  \includegraphics[width=8cm]{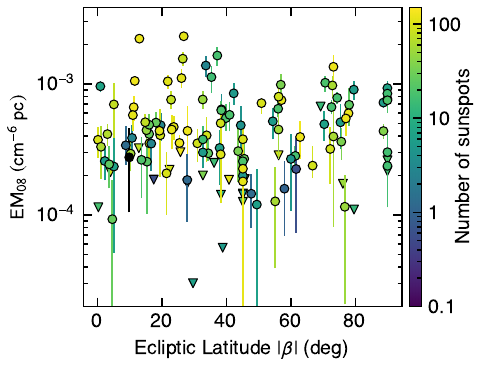}
 \end{center}
 \caption{\color{black} EM$_{08}$ plotted against the absolute value of the ecliptic latitude, $\beta$. The color scale indicates the 13-month averaged sunspot number.  }\label{fig:vsbeta}
\end{figure}

\color{black}
The heliospheric SWCX is expected to be stronger near the ecliptic plane \citep{Robertson03,Koutroumpa06}. 
Figure \ref{fig:vsbeta} shows EM$_{\rm 08}$ values of the 130 observations plotted against the absolute value of the ecliptic latitude, with the color scale of the 13 months averaged sunspot number. There is no clear dependence of EM$_{\rm 08}$ on the ecliptic latitude.
\color{black}

\subsection{Spectral fitting with the SWCX emission lines}\label{sec:sys}

\begin{figure*}
 \begin{center}
    \includegraphics[width=8cm]{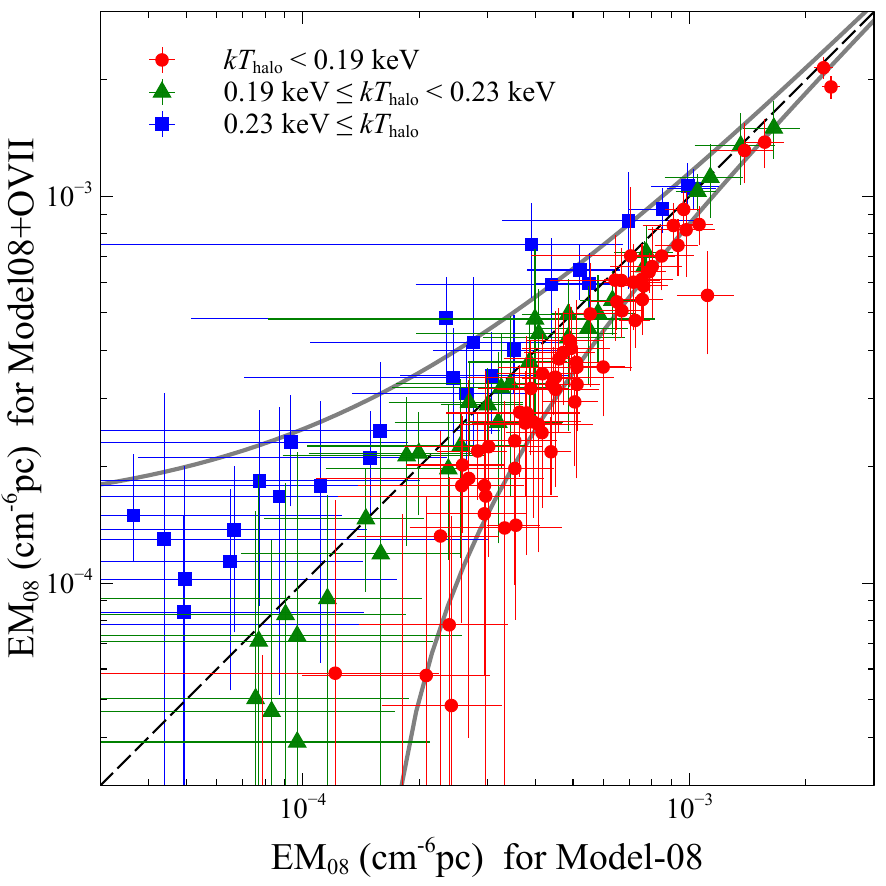}
     \includegraphics[width=8cm]{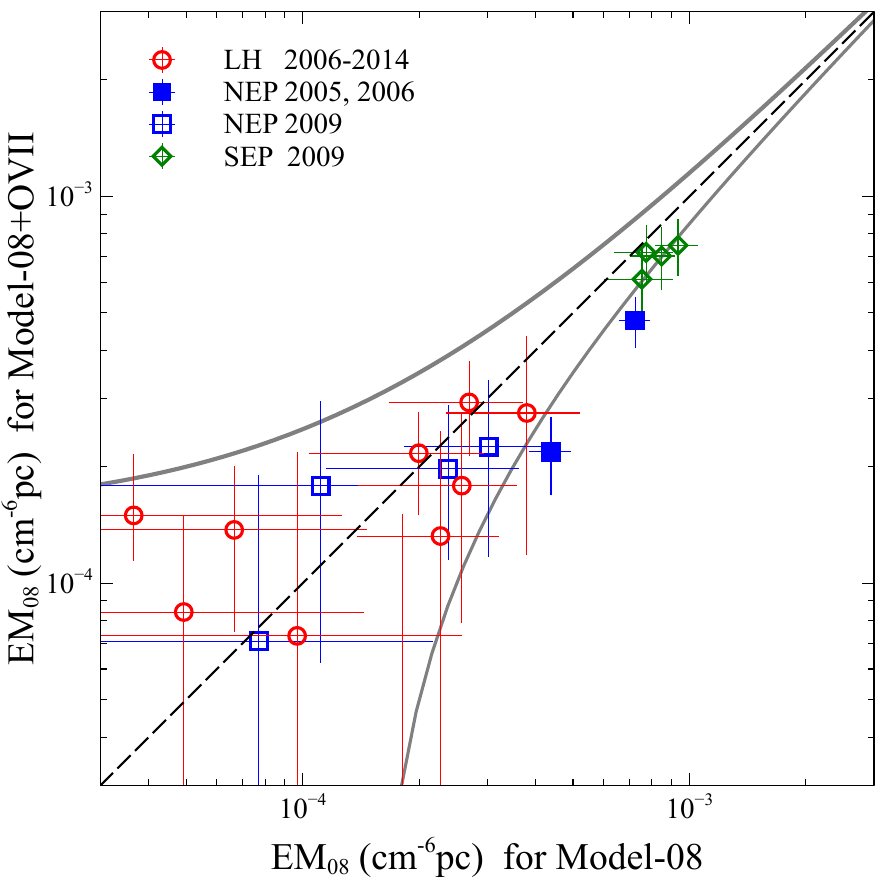}
  \includegraphics[width=8cm]{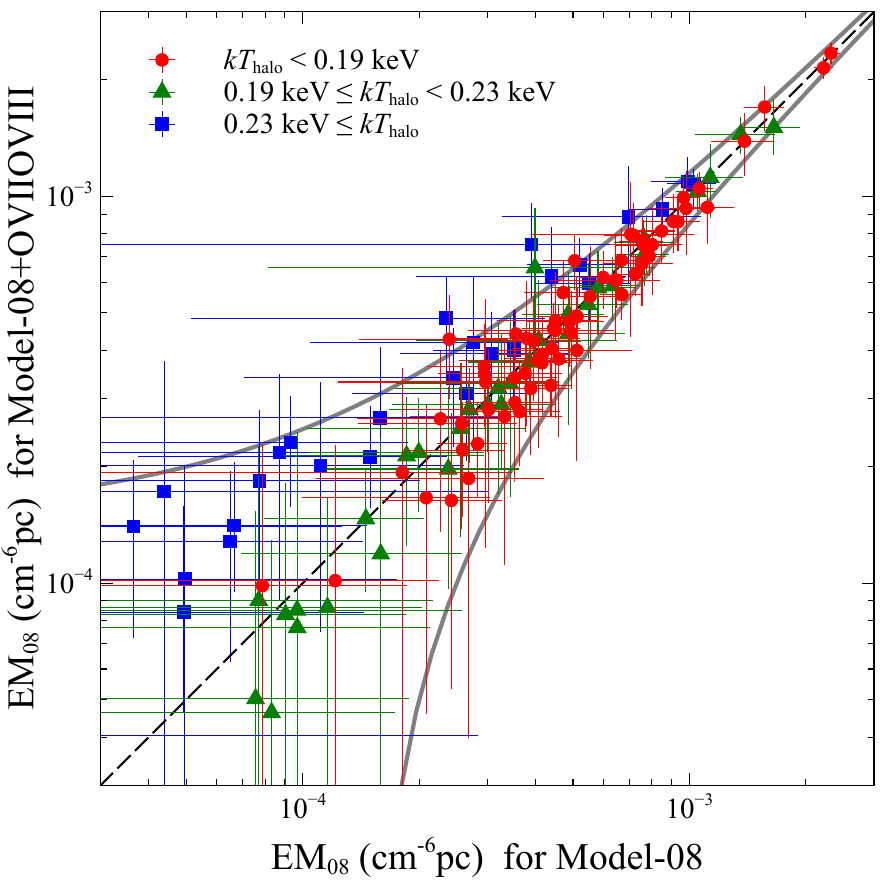}
   \includegraphics[width=8cm]{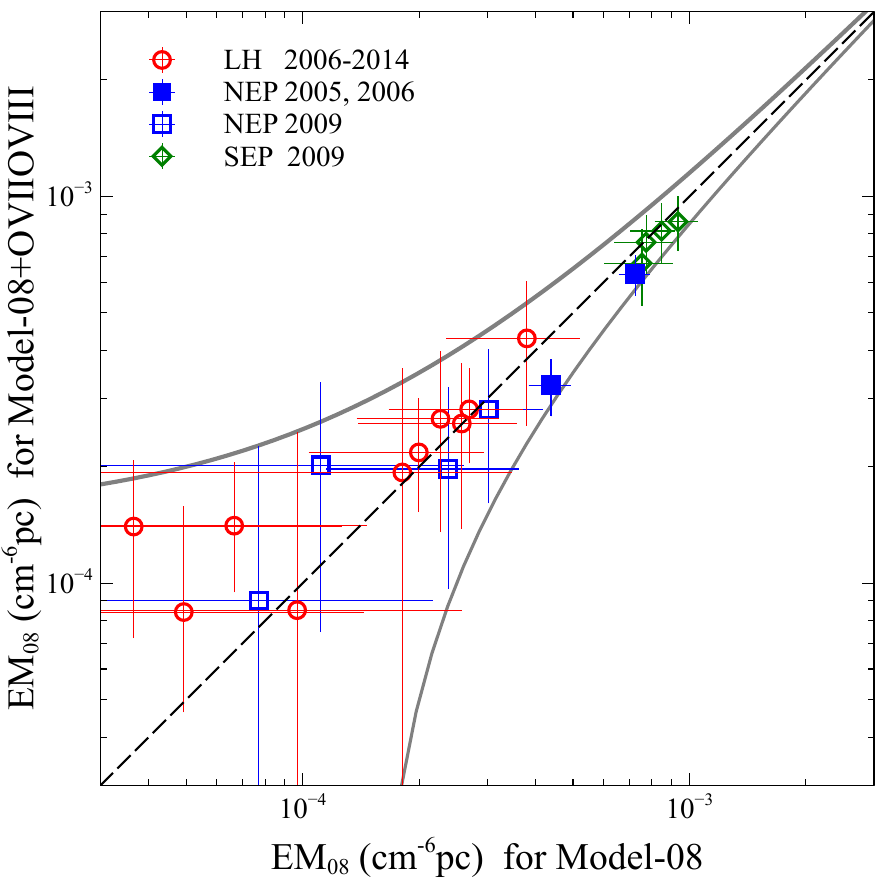}
\end{center}
\caption{(top left panel) EM$_{08}$ with Model-08+OVII plotted against those with Model-08 for  the data with $kT_{\rm halo}<0.19$ keV (filled circles), $0.19 \rm{keV}\le kT_{\rm halo}<0.23$ keV (filled triangles), and $kT_{\rm halo}>0.23$ keV (filled squares).
The dashed line is the line of equality, and the gray solid lines show the difference of 1.5$\times 10^{-4}~\rm{cm^{-6} pc}$.  
(top right panel) The same as the left panel but for LH (open circles), NEP (2005-2006; filled squares), NEP (2009; open squares), and SEP (open diamonds).
(bottom panels) Same as the top panels but for Model-08+OVII\color{black}O\color{black}VIII.
}\label{fig:model08vs022}
\end{figure*}

Paper-I reported that fitting the 2005-2009 data with Model-08 resulted in fairly uniform values of $kT_{\rm halo}$ with the median value of 0.22 keV.  In contrast, for the 2010-2015 data, Model-08 gives lower values of $kT_{\rm halo}$.
They concluded that during the solar maximum period, excess emissions of the O\,\emissiontype{VII} He$\alpha$ line possibly from the heliospheric SWCX cause an underestimation of  $kT_{\rm halo}$ and an overestimation of EM$_{\rm halo}$.
Therefore, we added a Gaussian with the fixed central energy of 0.56 keV for the O\,\emissiontype{VII} He$\alpha$ line to Model-08, fixing $kT_{\rm halo}$ at 0.22 keV, and re-fitted the spectra of the 130 observations. 
We will refer to this model as Model-08+OVII.
The differences in EM$_{08}$ between the two model fits are up to $\sim$1.5$\times 10^{-4}~\rm{cm^{-6}pc}$, which is much smaller than the scatter in EM$_{08}$ for Model-08.
As shown in the left panel of figure \ref{fig:model08vs022},
when $kT_{\rm halo}$ for Model-08 is higher than 0.23 keV, EM$_{08}$ for Model-08+OVII increases from that for Model-08.  With Model-08, an overestimation of $kT_{\rm halo}$  lead to an underestimation of EM$_{08}$.
In contrast, for the data with $kT_{\rm halo}<0.19$ keV,  EM$_{08}$ for Model-08+OVII decreases from that for Model-08.

\begin{figure*}
 \begin{center}
    \includegraphics[width=17cm]{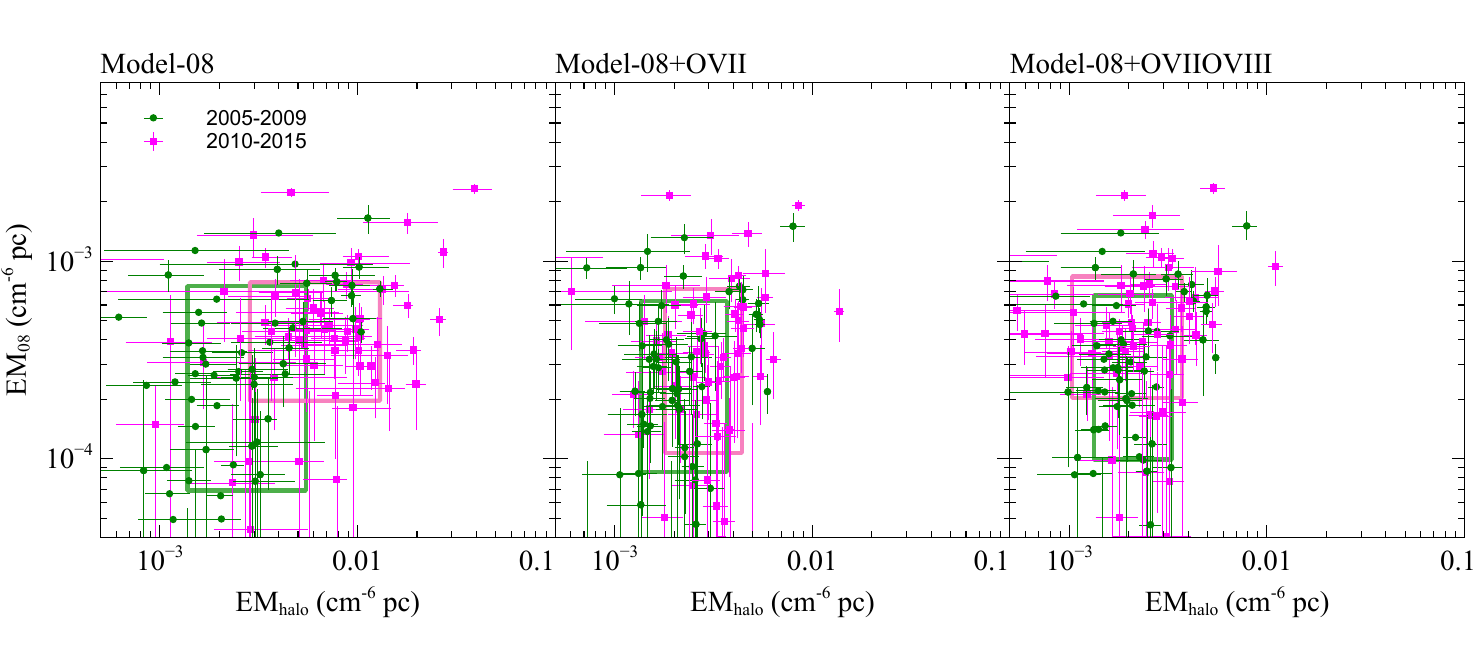}
\end{center}
\caption{EM$_{08}$ with Model-08 (left panel), with Model-08+OVII (middle panel), and with Model-08+OVII\color{black}O\color{black}VIII (right panel) plotted against \color{black} EM$_{\rm halo}$ \color{black}. Circles and squares correspond to the data taken before and after, respectively, the end of 2009.
The light magenta and green boxes show the 18th-84th percentile ranges for the 2005-2009 and 2010-2015 data, respectively.}\label{fig:mwhvsmwh2}
\end{figure*}

Our sample includes nine data toward Lockman Hole (LH) data obtained from 2006 to 2014, covering almost one solar cycle, with nearly the same lines of sight. Suzaku also observed the North Ecliptic Pole (NEP) in 2005, 2006, and four times in 2009 (whose pointings are 1$^\circ$.2 offset from those obtained in 2005 and 2006) and the South Ecliptic Pole (SEP) four times in 2009.
As shown in figure \ref{fig:model08vs022}, EM$_{08}$ with nearly the same pointing are more consistent with each other for Model-08+OVII than for Model-08.
In particular, the scatter in EM$_{08}$ is reduced for the LH data, while
the weighted averages of EM$_{08}$ from the two models are almost the same at $(1.6\pm 0.3)\times 10^{-4}~\rm{cm^{-6}pc}$. 
The left and middle panels of figure \ref{fig:mwhvsmwh2} show the plots EM$_{08}$ against EM$_{\rm halo}$ with Model-08 and Model-08+OVII, respectively.
The 16th-84th percentile ranges of EM$_{08}$ for the 2005-2009 and 2010-2015 data become more consistent with Model-08+OVII than those for Model-08 (table \ref{tab:em08median2}).
These consistencies in EM$_{08}$ with Model-08+OVII indicate that the 0.7-1 keV residual structures detected in some regions are unlikely to be caused by the remaining SWCX emissions in this energy band.
As reported by Paper-I, EM$_{\rm halo}$ with Model-08 depends on the solar activity (table \ref{tab:em08median2} and the left panel of figure \ref{fig:mwhvsmwh2}).  As plotted in the middle panel of figure \ref{fig:mwhvsmwh2}, with Model-08+OVII, the scatter in EM$_{\rm halo}$ is significantly reduced, and the 16th-84th percentile ranges for the two epochs become consistent.

\begin{table*}
            \tbl{Medians and 16th-84th percentile ranges of EM$_{\rm halo}$ and EM$_{08}$  }{
                      \begin{tabular}{cccccccccc}\hline 
                       &  &  &\multicolumn{2}{c}{EM$_{\rm halo}$} & \multicolumn{2}{c}{EM$_{\rm 08}$}\\
        model &   selection     & N$^*$ & median & 16th-84th percentile  & median & 16th-84th percentile\\
               &            &  &  $10^{-3}~\rm{cm^{-6}pc}$ & $10^{-3}~\rm{cm^{-6}pc}$ & $10^{-4}~\rm{cm^{-6}pc}$ & $10^{-4}~\rm{cm^{-6}pc}$\\\hline
Model-08 & 2005-2009 & 64 &  2.4 &  1.4-- 5.5 &   2.8 &  0.7-- 7.5\\
Model-08 & 2010-2015 & 66 &  6.8 &  2.9--13.0 &   4.2 &  2.0-- 7.8\\
Model-08+OVII & 2005-2009 & 64 &  2.0 &  1.4-- 3.7 &   2.9 &  0.9-- 6.3\\
Model-08+OVII & 2010-2015 & 66 &  3.0 &  1.8-- 4.4 &   3.4 &  1.1-- 7.2\\
Model-08+OVIIVIII &2005-2009 & 64 &  1.9 &  1.3-- 3.3 &   2.9 &  1.0-- 6.7\\
Model-08+OVIIVIII &2010-2015 & 66 &  2.4 &  1.0-- 3.7 &   4.5 &  2.0-- 8.4\\
	     \hline
            \end{tabular}}\label{tab:em08median2}
                \begin{tabnote}
                    \footnotesize
                    \footnotemark[$*$] Number of observations\\           
	      \end{tabnote}
                \end{table*}

With XMM-Newton observations, \citet{Qu22} reported that 
the  O\,\emissiontype{VIII} Ly$\alpha$ line emissions from the heliospheric SWCX also contaminate the spectra.
 Therefore, we added another Gaussian at 0.65 keV for O\,\emissiontype{VIII} Ly$\alpha$ to Model-08+OVII and re-fitted the spectra. 
Here we assumed that the normalization of the Gaussian at 0.65 keV is 20\% of that at 0.56 keV. This ratio is derived from the excess emission from the Lockman Hole data obtained in 2013 at the solar maximum compared to that obtained in 2009 at the solar minimum (Paper-I).
We will refer to this model as Model-08+OVII\color{black}O\color{black}VIII.
The differences in EM$_{08}$ are up to $\sim$1.5$\times 10^{-4}~\rm{cm^{-6}pc}$.
However, for the data with $kT_{\rm halo}<0.19$ keV for Model-08, Model-08+OVII\color{black}O\color{black}VIII gives almost the same EM$_{08}$ with Model-08.
  With this model, the median and the 16th-84th percentile range of EM$_{08}$  are similar to those for Model-08 (table \ref{tab:em08median2} and figure \ref{fig:mwhvsmwh2}).
  Although the error bars of the individual observations are relatively large, EM$_{08}$ with nearly the same lines of sight are more consistent for Model-08+OVII than those for Model-08+OVII\color{black}O\color{black}VIII.
As shown in figure \ref{fig:model08vs022}, 
when $kT_{\rm halo}$ for Model-08 is higher than 0.23 keV, EM$_{08}$ for Model-08+OVII\color{black}O\color{black}VIII also increases from that for Model-08.
For each observation, it is difficult to determine whether Model-08+OVII or Model-08+OVII\color{black}O\color{black}VIII is more appropriate.
The differences in EM$_{08}$ from the three models are up to 1.5$\times 10^{-4}~\rm{cm^{-6}pc}$.  
We will treat this value as a systematic error in EM$_{\rm 08}$.

\section{Discussion}

         We analyzed the Suzaku data of the 130 observations toward $75^\circ<l<285^\circ$ and $|b|>15^\circ$. 
                       Our sample does not include the eROSITA bubble \citep{erositabubble}, bright regions toward the north and south of the Galactic center.
                          With the standard soft X-ray background model consisting of LHB and MWH, residual structures remain in the spectra of some regions at 0.7--1 keV.
              For 56 out of the 130 observations, adding a spectral component with $kT\sim 0.8$ keV, which is much higher than the virial temperature of the Milky Way, gives better fits than the standard soft X-ray background model.  \color{black} This component tends to be brighter toward the lower Galactic latitude\color{black}.
              A similar supervirial temperature component has been detected with the HaloSat observations of 85\% of the field \color{black} at $|b|>30^\circ$\color{black} \citep{Halo22}.
           The EM$_{08}$ around the eROSITA bubble observed with Suzaku tends to be higher than the other regions \citep{Gupta2022}.
Excluding the eROSITA bubble, the detection rate of EM$_{08}$ with HaloSat becomes closer to ours.
\color{black}Beyond 100$^\circ$ from the Galactic center, the median value of EM$_{08}$ with Halosat  is  $\sim 10^{-3}~\rm{cm^{-6}pc}$.  This value corresponds to  $\sim 4 \times 10^{-4}~\rm{cm^{-6}pc}$ for the solar abundance table we use and is consistent with our median value. \color{black}

              Paper-I reported that $kT_{\rm halo}$ and EM$_{\rm halo}$ become fairly uniform toward the Galactic anti-center sky and at high Galactic latitude and suggested that the virial temperature plasma fills the Milky Way halo.
              As shown in figure \ref{fig:mwhvsmwh2}, the scatter in EM$_{08}$ is significantly larger than that in EM$_{\rm halo}$. Therefore, 
it is reasonable to assume that the MWH component is smoothly distributed in the halo and that the 0.8 keV component has a different origin than the MWH component.

\subsection{Contamination of unresolved stars}
 
 The spectra of the 0.8 keV component resemble those of stars.
 \citet{Misui09} found a CIE component with a temperature of $\sim$0.8 keV in the direction of 
 ($l,b$)=(235$^\circ$, $0^\circ$) with Suzaku.
 The spectral shape and emission measure of this component are consistent
 with a sum of unresolved faint dM stars.
  \citet{Yoshino09} also detected a similar spectral shape component from a Suzaku observation toward $b=\sim 10^\circ$. However, its emission measure 
   is a factor of 5 higher than the predicted value of the contribution from dM stars, taking into account the expected decrease in the stellar surface brightness toward the higher Galactic latitudes.
  The emission measure of this $b=10^\circ$ component is similar to those of our sample at $|b|=15^\circ\sim 20^\circ$.
    
The total resolved stellar flux toward the Chandra Deep Field South ($l=224^\circ, b=-54^\circ$) down to 5$\times 10^{-18}~ \rm{erg s^{-1}cm^{-2}}$ is about a few percent of that from AGNs in the 0.5--2.0 keV energy band \citep{Chandra12}.
This stellar contribution is equivalent to EM$_{08}\sim 10^{-4}~\rm{cm^{-6} pc}$.
The weighted average of EM$_{08}$ from the nine LH observations ($l=150^\circ, b=53^\circ$) is $(1.6 \pm 0.3)\times 10^{-4}~\rm{cm^{-6} ~pc}$ for both Model-08 and Model-08+OVII. This value is marginally consistent with the stellar contribution from the Chandra observations with similar $|l|$ and $|b|$.
The EM$_{08}$ of the LH observations is one of the lowest among the 130 observations.
A significant fraction of the observations at $|b|>50^\circ$ show much higher EM$_{08}$.
At $|b| = 30^\circ$, the integrated star count with 2MASS is a factor of  $\sim $two higher than that at $|b|\sim 50^\circ$ \citep{2MASS2011}.
Assuming that the stellar contribution to the 0.8 keV component is proportional to the integrated star counts, EM$_{08}$ at $|b|\sim 30^\circ$ from stars is about a few times $10^{-4}~\rm{cm^{-6} ~pc}$, which is lower than the median value of EM$_{08}$ at $35^\circ<|b|< 50^\circ$. 

If the excitation radius for bright sources is small, their scattered photons can contaminate the spectra. 
For example, there are two bright stars in the FOV of the 2005 LH observation, which is offset by 0$^\circ$.5 from the other LH observations.
When the excitation radius is 1$'$, the best-fit value of EM$_{08}$ for the 2005 data, $6\times 10^{-4}~\rm{cm^{-6} ~pc}$, is significantly higher than those for the other LH observations.
In this paper, the excitation radius of $2'.5\sim 3'$ is adopted for these stars. Then, the derived EM$_{08}$, $(2.7\pm 1.0) \times 10^{-4}~\rm{cm^{-6} ~pc}$, is consistent with the weighted average value of EM$_{08}$ of the other LH observations.
Finally, if the contribution of the faint stars is significant, we expect a smoother distribution of EM$_{08}$.
 Thus, the stellar contribution to the 0.8 keV component is relatively small.

\subsection{Contamination of background groups of galaxies}\label{sec:group}

The spectrum of the 0.8 keV component also resembles those from intracluster medium (ICM) in groups of galaxies.
 If the hot gas has an extragalactic origin, we expect a uniform distribution of the hot gas blob without dependence on the Galactic latitude and longitude.
Therefore, the background groups and clusters of galaxies may not be the primary origin of the 0.8 keV component.
However, regions with the brightest EM$_{08}$, especially at high latitudes, may be related to the ICM emission from galaxy groups.
The 0.8 keV galaxy groups typically have $r_{500}$ of $\sim$ 0.4 Mpc \citep{Sun09,group17}.
Here, $r_{500}$  is the radius of a sphere whose mean enclosed density is 500 times the critical density of the Universe.
At $z=0.03$, this radius is $\sim$11 arcmin.
Since the 0.8 keV component extends across the FOV of the XIS detectors, 
we searched for concentrations of galaxies at $z<0.03$ in each FOV with 
$\Delta C<-9$. 
There are two Hickson compact groups \citep{HCG}, HCG 16 at $z$=0.012,  HCG 96 at $z$=0.029, and two groups at $z$=0.029 and 0.015 \citep{Group2,Group3} in the FOVs of observations with ObsI{\color{black}D}=709009010, 708023010, 703016010, 708002010, respectively.

\subsection{Modeling of the spatial distribution of the 0.8 keV component}
\label{sec:spmodel}

            \begin{figure*}
 \begin{center}
\includegraphics[width=16cm]{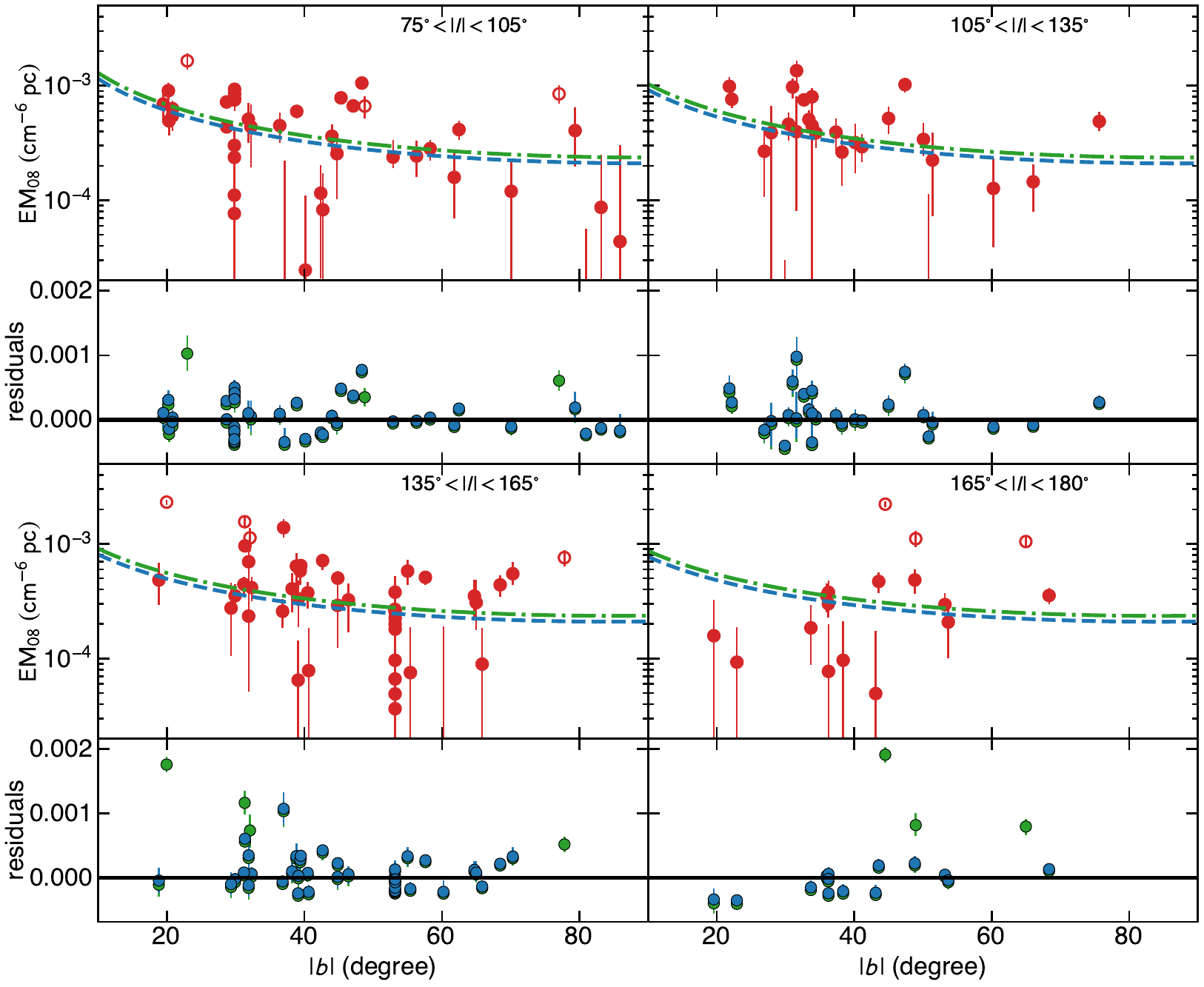}
 \end{center}
\caption{EM$_{08}$ with Model-08, plotted against the Galactic latitude, $|b|$, separated into four $|l|$ ranges. The dot-dash and the dashed lines show the best-fit disk-like morphology models ($R_0=3$ kpc and $z_0=0.3$ kpc) for all data (filled and open circles for the upper panels and open circles for the bottom panels) and those excluding the regions toward Orion-Eridanus superbubble, TW Hya, and possible candidates of groups of galaxies (plotted as open circles), respectively. The bottom panels show the residuals.}\label{fig:em08vsgalactic}
\end{figure*}

Figure \ref{fig:em08vsgalactic} shows EM$_{08}$ plotted against $|b|$, separated into four $|l|$ ranges.  The scatter in EM$_{08}$ toward the Galactic anticenter ($165^\circ <|l| < 180^\circ$) is relatively small, except for one possible group candidate (ObsID=709009010, $l=174^\circ, b=-65^\circ$, section \ref{sec:group}) and the two regions toward the Orion-Eridanus superbubble within this $|l|$ range.
We model the spatial distribution of the 0.8 keV component, assuming that it is a CIE plasma.
The possible dependence of EM$_{08}$ on $|b|$ suggests that the emissivity increases toward the disk.
Therefore, we compare the obtained EM$_{08}$ distribution with the following disk-like gas model,

\begin{equation}
n_{\rm e}=n_{\rm e0}\exp\left(-\frac{R}{R_0}\right)\exp\left(-\frac{z}{z_0}\right)\label{eq:disk}
	\end{equation}

Here, $n_{\rm e}$ is the electron number density, $n_{\rm e0}$ is that at the Galactic center, $R$ is the distance from the Galactic center projected onto
the Galactic plane, $z$ is the vertical height, $R_0$ and $z_0$ are 
the scale length and scale height, respectively.
Assuming the distance from the Solar System to the Galactic Center to be 8 kpc, we fit the EM$_{08}$ distribution of the 130 observations by integrating the equation \ref{eq:disk}.
Since it is difficult to constrain $R_0$ and $z_0$, we fixed $R_0$ at 3 kpc and $z_0$ at 0.3 kpc, which are close to those of the thin disk \citep{2MASS2011, Cautun20}, and estimated ${n_{\rm e}}_0f_{08}^{1/2}$. Here, $f_{08}$ is the volume filling factor of the 0.8 keV component.
The fit is unacceptable with $\chi^2$/d.o.f=1372/129, as shown in table \ref{tab:density model} and figure \ref{fig:em08vsgalactic}.
We then excluded the five regions toward Orion-Eridanus superbubble,   TW Hydrae,  
and the four group candidates (see section \ref{sec:group} in detail).
The fit is still not acceptable with $\chi^2$/d.o.f=744/119.
To account for possible systematic uncertainties caused by contamination of the SWCX (section \ref{sec:sys}), we added systematic errors of $1.5\times 10^{-4}~\rm{cm^{-6}pc}$ to each EM$_{08}$.
This reduced $\chi^2$ significantly reduced to $\chi^2$/d.o.f=197/119.
As summarized in table \ref{tab:density model}, ${n_{\rm e}}_0f_{08}^{1/2}$ does not depend on the sample selection and systematic uncertainties.
Assuming $R_0=7$ kpc, which is the value adopted for the MWH component in Paper-I, 
${n_{\rm e0}f_{08}}^{1/2}$ decreases by about a factor of 5. However, around the solar system, or at $R=8$ kpc and $z=0$ kpc\color{black}, \color{black}  the two $R_0$ cases yield almost the identical values of ${n_{\rm e}}f_{08}^{1/2}=(1.3\--1.4)\times 10^{-3}~\rm{cm^{-3}}$. 
If we allow $z_0$ to vary,  fixing $R_0$ at 7 kpc, we get $z_0=1.9\pm 1.2$ kpc.
Although ${n_{\rm e0}f_{08}}^{1/2}$ depends strongly on $R_0$ and $z_0$, at $R=8$ kpc, and $z=0.3$ kpc, 
all the five fits give almost the same value of ${n_{\rm e}}f_{08}^{1/2}=0.5\times 10^{-3}~\rm{cm^{-3}}$.
 
 \begin{table*}
\tbl{Fitting results of the density distribution models}{
\begin{tabular}{clrrrcccr}\hline
selection                & $N^*$   & ${n_{\rm e0}f_{08}^{1/2}}^{\dagger}$ & $R_0$ & ${z_0}$ &
 $\chi^2$/d.o.f \\
                &    &  ($10^{-3} {\rm cm}^{-3}$) &  (kpc)  &  (kpc) &\\
                \hline
all &    130 & 19$\pm$1 & 3.0 (fix) & 0.3 (fix) & 1372/129 \\
all - bright regions$^\ddagger$ & 120 & 18$\pm$1 & 3.0 (fix) & 0.3 (fix) & 744/119 \\
all - bright regions$^\ddagger$$^\S$ & 120 & 20$\pm$1 & 3.0 (fix) & 0.3 (fix) &  197/119 \\
all - bright regions$^\ddagger$$^\S$ & 120 & 4.2$\pm$0.1 & 7.0 (fix) & 0.3 (fix) & 199/119\\
all - bright regions$^\ddagger$$^\S$ & 120 & 1.8$\pm$0.5 & 7.0 (fix) & 1.9$\pm$1.2 & 194/118\\
 \hline
\end{tabular}}\label{tab:density model}
 \begin{tabnote}
                    \footnotesize
                    \footnotemark[$*$] Number of observations\\
	               \footnotemark[$\dagger$] The central electron density ($n_{e0}$)  of the disk model, assuming the metal abundance of 1 solar. $f_{08}$ is the volume filling factor of the 0.8 keV component.\\
	               \footnotemark[$\ddagger$] The five regions toward the Orion-Eridanus superbubble, one toward TW Hya, and four toward group candidates were excluded from the sample.\\
	               \footnotemark[$\S$] A systematic error of $1.5\times 10^{-4}~\rm{cm^{-6}pc}$ is added in quadrature to each EM$_{08}$.\\
	            	                 \end{tabnote}
\end{table*}

\subsection{Physical properties of the 0.8 keV component}

The median values of the emission measures of the 0.8 keV and MWH components for the 2005-2009 data are (2.8--2.9)$\times 10^{-4}~\rm{cm^{-6} pc}$ and (2.0--2.4)$\times 10^{-3}~\rm{cm^{-6} pc}$, respectively, with the three models, 
Model-08, Model08+OVII, and Model-08+OVIIOVIII (table \ref{tab:em08median2}).
If the two components follow the same spatial distribution and are in pressure equilibrium,  $f_{\rm halo}$ is a factor of $\sim$ 1.5--2 smaller than $f_{08}$.  Here,  $f_{\rm halo}$ is the volume filling factor of the MWH component, and we assume the temperature of the two components at 0.8 keV and 0.22 keV.
If this is true, then the 0.8 keV component fills a larger spatial volume than the MWH component, and we expect a smoother distribution for the 0.8 keV component than for the MWH component.
However, the EM$_{\rm halo}$ is quite uniform compared with EM$_{08}$ (figure \ref{fig:mwhvsmwh2}).
Paper-I reported that the emission distribution of the MWH component at $|l|>105^\circ$ is well represented by a sum of disk-like and spherical morphology components.
If a significant fraction of the MWH component comes from the latter, the scatter in EM$_{\rm halo}$ may be small.

The scatter in EM$_{08}$  suggests a relatively small value of $f_{08}$.
There are other ISM components in the Milky Way.
At the midplane, most of the volume is filled with neutral medium: assuming turbulent pressure equilibrium, its volume filling factor is estimated to be about 0.6, and that for the hot ionized medium is only about 0.2 \citep{Kalberla09}.
They also estimated the vertical structure of these media: at $z=0.5$ kpc, the volume filling factor of the hot gas reaches 0.3--0.4.
Above the midplane, the volume filling factor of the warm ionized medium (WIM), or H$\alpha$-emitting gas, increases, and at $z\sim 0.5-1$ kpc, peaking at about 0.3.
The pressure of WIM is estimated to be $p/k\sim 5000 ~\rm{cm^{-3}K}$ and a few thousand $\rm{cm^{-3}K}$ at the midplane and at $z=0.3$ kpc, respectively \citep{Gaensler2008}.
Above this height, a hot ionized medium is expected to fill more than half of the volume.

Adopting the best-fit disk-like and spherical morphology model for the MWH component in Paper-I and the disk-like model for the 0.8 keV component, 
we can estimate the electron density, cooling time, and pressure. 
At $R=8~{\rm kpc}$ and $z=0.3~{\rm kpc}$, $n_{\rm e}f_{08}^{1/2}=0.5\times 10^{-3}~\rm{cm^{-3}}$ 
and  $n_{\rm e, 022}f_{\rm halo}^{1/2}=10^{-3}~\rm{cm^{-3}}$. Here, $n_{\rm e, 022}$ is the electron density of the MWH component.
Then, the radiative cooling time of the 0.8 keV component is 7$f_{08}^{1/2}$ Gyr.
The pressures, $p$, of the 0.8 keV and MWH components are $p/k\sim 9000~f_{08}^{-1/2}\rm{cm^{-3}K}$ and $5000~f_{\rm halo}^{-1/2}\rm{cm^{-3}K}$, respectively. 
If these two components are in pressure equilibrium, $f_{08}$ is a factor of three larger than $f_{\rm halo}$.
 Assuming that the volume filling factor of the hot gas (MWH and 0.8 keV) is 0.25, the hot gas pressure reaches 20000~$\rm{cm^{-3}K}$, which is much higher than the thermal pressure of WIM.
Considering that the temperature of the 0.8 keV component is significantly higher than the virial temperature and its long cooling time and high pressure, it can escape from the Galactic disk.

\subsection{The Orion-Eridanus superbubble}\label{sec:ori}

Among the observations with the largest EM$_{08}$, five are 
toward  the Orion-Eridanus superbubble around 
($l, b$)$\sim$ (200$^\circ$, -40$^\circ)$.
This superbubble is a nearby ($\sim 400$ pc) and well-known feature of the soft X-ray background spanning 20$^\circ\times 45^\circ$ on the sky (e.g. \cite{Reynolds79}, \cite{Burrows1993}, \cite{Snowden1995}), 
and possibly a large cavity extending from the Ori OB1 association.
\citet{Burrows1993} reported that a 0.1--0.2 keV plasma fills the interior of the superbubble, which extends $\sim 200$ pc from the Galactic disk. 
The cavity is thought to be formed from SN remnants, UV radiation, or stellar wind from the Orion OB1 association.
The enhancement of the 0.8 keV component toward the Orion-Eridanus superbubble indicates that a hotter plasma is also likely to exist in the bubble.
With HaloSat observations, \citet{Fuller2023} detected a hot plasma with a temperature of 0.79 $\pm$ 0.12 keV.
The pressure of this component is estimated to be (3--5)$\times 10^4~\rm{cm^{-3}K}$, which is a factor of two higher than that of the 0.8 keV component at $R=8$ kpc and $z_0$=0.3 kpc.
Numerical simulations expect that young star clusters form superbubbles whose evolution is quite different from those of each SN (e.g.~\cite{Keller2014}, \cite{Sharma2014}).
These simulations suggest that SNe in star clusters may heat some gas to $\sim$ 1 keV, whose cooling time is much longer than 10$^8$ yr.
If so, the Orion OB1 association may have produced hot plasma with a temperature of $\sim$ 1 keV. 
 

  \subsection{The origin of the 0.8 keV component}

Strong emissions with a temperature of about 1.3 keV are seen in the Galactic ridge spectra at $2^\circ<|l|<30^\circ$ and $|b|<1^\circ$ (\cite{Uchiyama13}).
The estimated electron density of this ridge component is about 2 $\times 10^{-3}~\rm{cm^{-3}}$, which is close to $n_{\rm e0}$ of the 0.8 keV component if it has a relatively large $R_0$, about 7 kpc.
In spiral galaxies, the temperature component around 0.8 keV is sometimes detected in addition to the 0.2--0.3 keV component (e.g. \cite{Mineo12}, \cite{M83XMM20}).
The estimated $n_{\rm{e}}$ of the 0.8 keV component in our analysis is comparable to that for other spiral galaxies such as M83 (\cite{M83XMM20}).
These results suggest that the 0.8 keV component is common in spiral galaxies.

The enhancement of the 0.8 keV component around the Orion-Eridanus superbubble
suggests that SNe or stellar wind can heat the gas to $\sim 1 $ keV.
Stars are formed in clusters, and SNe create superbubbles that are more effective at heating the interstellar medium than each SN.
The temperature of the hot gas in such superbubbles reaches 1 keV, with a relatively long cooling time
\citep{Keller2014,Keller2016}.
This temperature is higher than the virial temperature of the Milky Way.
Therefore, bubbles can leave the stellar disk and cause galactic fountains.

\color{black}
Based on the detection of 0.5--1 keV component, the influence of the stellar feedback has been discussed in previous studies (e.g. \cite{Nakashima18},\cite{Halo22}, \cite{Ponti22}). Then, we expect that EM$_{08}$ increase toward the Galactic disk, where star formations occur. Our dataset, which focuses on $|b|>15^\circ$, indicates a dependence on $|b|$. 
Although the error is large, the best-fit scale height for the 0.8 keV component, $z_0=1.9\pm 1.2$ kpc, is higher than the scale height of the stellar disk.
These results \color{black} may be related to the recent feedback from the Galactic disk or the Galactic fountains.
 The scatter in EM$_{08}$ is consistent with stellar feedback.

Paper-I reported that with Model-08, in the stacked spectra of the 130 observations, there is a residual at the line energies of the Ne Ly$\alpha$ line and the Mg He$\alpha$ line.
 If SNe produce a hot gas in a star-forming region, we expect an enhancement of the abundances of the $\alpha$-elements \citep{Nomoto2013}.
When we allow $kT_{\rm 08}$ to vary, with Model-08-kTfree, some regions have higher temperatures around 1.3 keV.
Such high-temperature regions may contribute to these Ne and Mg lines.
If the hot gas is multiphase, as expected from numerical simulations (e.g., \cite{Keller2014}), it is a challege to degenerate the temperature structure and abundance pattern with the energy resolution of the CCD detectors.
With {\it XRISM}, the successor to the {\it Hitomi} mission,  observations of spiral galaxies will provide a valuable opportunity to access the temperature structure and reliable abundance pattern \citep{XRISM2020}.

 \section{Summary and Conclusion}
 
 We analyzed XIS data from 130 Suzaku observations at 75$^\circ<l<285^\circ$
 and $|b|>15^\circ$.
 We fitted the spectra with a standard soft X-ray background model consisting
 of the LHB and MWH. Then, there are residual structures at 0.7--1 keV in 
 some regions.
 Including an additional hot CIE component with a temperature of $\sim$0.8 keV
 gives $\Delta C<-9$ for 56 observations out of 130.
EM$_{08}$ show a significant scatter, with the median value of 4$\times 10^{-4}~ \rm{cm^{-6} pc}$ and the 16th-84th percentile range of (1--8)$\times 10^{-4}~ \rm{cm^{-6} pc}$.
 The data were obtained from 2005 to 2015, covering nearly one solar cycle,
 and are suitable for studying the contamination by emissions from SWCX.
 Paper-I reported that the contamination of the O\,\emissiontype{VII} He$\alpha$ line,
 possibly from the heliospheric SWCX, sometimes leads to an underestimation of the temperature of the MWH component. From the spectral fits with the O\,\emissiontype{VII} He$\alpha$ and O\,\emissiontype{VII} Ly$\alpha$ lines with a fixed MWH temperature of 0.22 keV, we conclude that the contamination of the SWCX causes some systematic differences up to $1.5\times 10^{-4}~ \rm{cm^{-6} pc}$ in EM$_{08}$.
Regions toward the Orion-Eridanus superbubble have the highest emission measures of the 0.8 keV component. 
While the scatter is large,  the emission measures tend to be higher toward the lower Galactic latitude, indicating the Galactic origin of the 0.8 keV component.
The median value of EM$_{08}$ is significantly higher than the expected stellar contribution. The 0.8 keV component is likely produced by
supernovae in the Milky Way disk, possibly related to galactic fountains.

\bigskip
\begin{ack}
	We acknowledge all members who contributed to the Suzaku project.
We acknowledge support from JSPS/MEXT KAKENHI grant numbers,
16K05300 (KM), 17K05393 (KS), 18H01260 (NY).
	
\end{ack}

\scriptsize

\end{document}